\documentclass[jgrga]{agu2001}
\usepackage{graphicx}
\authorrunninghead{LIU ET AL.}

\titlerunninghead{Plasma Depletion and Mirror Waves Ahead of ICMEs}

\authoraddr{Y. Liu,
Kavli Institute for Astrophysics and Space Research, Massachusetts
Institute of Technology, Room 37-676a, Cambridge, MA 02139, USA.
(liuxying@space.mit.edu)}

\begin{document}

%
%
%
\setkeys{Gin}{draft=false}
%
%

%
%

\title{Plasma Depletion and Mirror Waves Ahead of Interplanetary Coronal Mass Ejections}


 \authors{Y. Liu\altaffilmark{1},
 J. D. Richardson\altaffilmark{1},
 J. W. Belcher\altaffilmark{1},
 J. C. Kasper\altaffilmark{1},
 and R. M. Skoug\altaffilmark{2}}

 \altaffiltext{1}{Kavli Institute for Astrophysics and Space Research,
 Massachusetts Institute of Technology, Cambridge, MA, USA.}

 \altaffiltext{2}{Space Science and Applications, Los Alamos National
 Laboratory, Los Alamos, NM, USA.}

%
%

\begin{abstract}
We find that the sheath regions between fast interplanetary coronal
mass ejections (ICMEs) and their preceding shocks are often
characterized by plasma depletion and mirror wave structures,
analogous to planetary magnetosheaths. A case study of these
signatures in the sheath of a magnetic cloud (MC) shows that a
plasma depletion layer (PDL) coincides with magnetic field draping
around the MC. In the same event, we observe an enhanced thermal
anisotropy and plasma beta as well as anti-correlated density and
magnetic fluctuations which are signatures of mirror mode waves. We
perform a superposed epoch analysis of ACE and Wind plasma and
magnetic field data from different classes of ICMEs to illuminate
the general properties of these regions. For MCs preceded by shocks,
the sheaths have a PDL with an average duration of 6 hours
(corresponding to a spatial span of about 0.07 AU) and a proton
temperature anisotropy ${T_{\perp p}\over T_{\parallel p}}\simeq
1.2$ -1.3, and are marginally unstable to the mirror instability.
For ICMEs with preceding shocks which are not MCs, plasma depletion
and mirror waves are also present but at a reduced level. ICMEs
without shocks are not associated with these features. The
differences between the three ICME categories imply that these
features depend on the ICME geometry and the extent of upstream
solar wind compression by the ICMEs. We discuss the implications of
these features for a variety of crucial physical processes including
magnetic reconnection, formation of magnetic holes and energetic
particle modulation in the solar wind.
\end{abstract}

%
%

%

\begin{article}

\section{Introduction}
Plasma depletion layers (PDLs) and mirror mode waves are common
features of planetary dayside magnetosheaths [e.g., Kaufmann et al.,
1970; Crooker et al., 1979; Hammond et al., 1993; Violante et al.,
1995; Hill et al., 1995]. The physical explanation for the plasma
depletion is that the plasma is squeezed out of the local noon
region as the magnetic field drapes around the magnetosphere [Zwan
and Wolf, 1976]. Combining this model with the double-adiabatic
equations derived by Chew et al. [1956], Crooker and Siscoe [1977]
predict that a temperature anisotropy of $T_{\perp}>T_{\parallel}$
is a direct consequence of the magnetic field line draping and
plasma depletion, where $\perp$ and $\parallel$ denote the
directions perpendicular and parallel to the magnetic field. When
the temperature anisotropy exceeds the instability threshold, mirror
mode waves are generated.

\begin{figure}
\noindent\includegraphics[width=20pc]{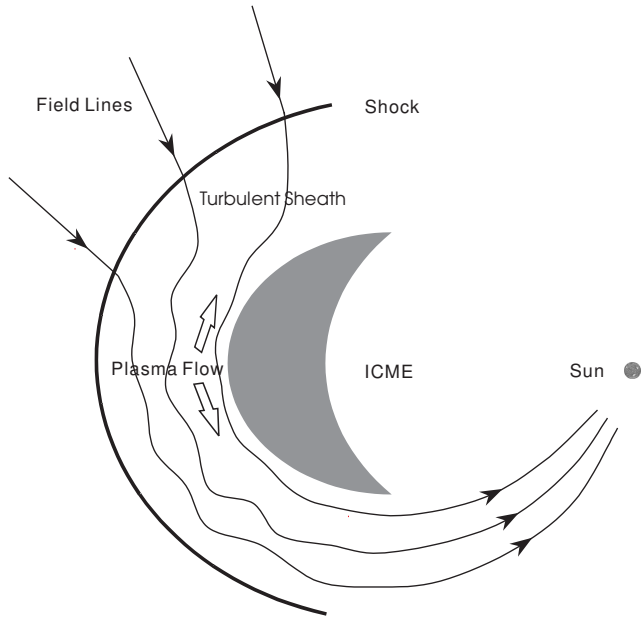} \caption{Schematic
diagram of the turbulent sheath between an ICME and the preceding
shock in the solar equatorial plane, illustrating the field line
draping and consequent plasma flow.}
\end{figure}

Magnetic field draping may occur in front of an obstacle in the
solar wind if the magnetized plasma cannot penetrate the obstacle.
Coronal mass ejections (CMEs), large-scale magnetic structures
expelled from the Sun, form large obstacles called interplanetary
coronal mass ejections (ICMEs) as they propagate into the solar
wind. The ambient magnetic field may be draped around fast ICMEs,
similar to the case of planetary magnetosheaths [Gosling and
McComas, 1987; McComas et al., 1988]. Figure~1 shows an idealized
sketch of this field line draping. An ICME with its axis lying in
the solar equatorial plane creates a preceding shock if its speed
relative to the ambient solar wind exceeds the fast-mode speed. Note
that the ICME may still remain magnetically connected to the Sun. In
front of the shock, the ambient magnetic field has the form of a
Parker spiral. Between the shock and ICME is the sheath region where
the magnetic field becomes stretched and turbulent. The shocked
solar wind plasma is compressed in the direction perpendicular to
the magnetic field, which may cause the plasma to flow along the
draped field lines and result in formation of a PDL. The plasma
temperature is enhanced in the perpendicular direction by the
compression and depressed along the magnetic field by the parallel
flow. As a result, anisotropic ion distributions are produced [e.g.,
Crooker and Siscoe, 1977]. When the threshold condition
\begin{equation}
{T_{\perp p} \over T_{\parallel p}} - 1 > {1 \over \beta_{\perp p}}
\end{equation}
is satisfied, the plasma is unstable to the mirror mode instability
[Chandrasekhar et al., 1958]. The perpendicular plasma beta of
protons is defined as $\beta_{\perp p} = {n_pk_{\rm B}T_{\perp p}
\over B^2/2\mu_0}$ with $n_p$, $k_{\rm B}$, $B$ and $\mu_0$
representing the proton density, Boltzmann constant, magnetic field
strength and permeability of free space, respectively. This picture
should apply to ICMEs with any axis orientation. Therefore, PDLs and
mirror mode waves are expected to form ahead of fast ICMEs. The
presence and properties of these features may depend on the geometry
of the ejecta and how much the magnetic field is draped and
compressed. Unlike the case of planetary magnetosheaths, ICMEs
expand in the solar wind and the spatial span of their sheath
regions increases since the preceding shock usually moves faster
than the ejecta.

A possible PDL lasting about 3 hours in front of an ICME was
identified and investigated by Farrugia et al. [1997], based on an
ideal magnetohydrodynamic (MHD) theory developed by Erkaev et al.
[1995]. Mirror mode structures which appear to be ahead of an ICME
were also found in the solar wind [Tsurutani et al., 1992]. In this
paper, we use a case study and statistical analysis to investigate
plasma depletion and mirror waves associated with ICMEs. Section 2
describes the observational data and data analysis. Section 3 shows
an example of a PDL and mirror waves in the sheath region of an
ICME. A superposed epoch analysis (SEA) of ICMEs, presented in
section 4, gives evidence that these features are usually observed
ahead of ICMEs preceded by shocks. We close by summarizing and
discussing the results in section 5 and test the SEA in the
Appendix. This paper provides the first consistent view of plasma
depletion and mirror waves in the environment of ICMEs.

\section{Observations and Data Reduction}
We use solar wind plasma and magnetic field observations at 1 AU
from ACE and Wind. Particularly important are the temperature
anisotropy data which are needed to identify mirror mode waves. The
ACE and Wind plasma teams use two different algorithms to calculate
$T_{\perp}$ and $T_{\parallel}$. The Wind team fits the measured ion
velocity distributions with a convecting bi-Maxwellian function,
using the magnetic field data to determine the parallel and
perpendicular directions [e.g., Kasper et al., 2002]. The accuracy
of their thermal anisotropies may be affected by angular
fluctuations of the magnetic field; spectra with angular
fluctuations over $15^{\circ}$ are rejected. The ACE team calculates
the second moment integrals of the measured distributions to obtain
the temperature anisotropy. The resulting temperature matrix is then
rotated into a field-aligned frame, which gives two perpendicular
and one parallel temperature [e.g., Gary et al., 2001]. If the
plasma is gyrotropic (i.e., particle gyration in a plane
perpendicular to the magnetic field has no preferred direction), the
two perpendicular temperatures should be the same; we only include
the data for which the ratio of the two perpendicular temperatures
is less than 1.3.

We use mainly ICMEs from the list of Liu et al. [2005] which are
observed at both ACE and Wind. Events with $T_p/T_{ex} \leq 0.5$ and
$n_{\alpha}/n_p \geq 8\%$ were qualified as ICMEs, where $T_p$,
$T_{ex}$ and $n_{\alpha}$ are the proton temperature, expected
proton temperature and alpha density separately. The expected
temperature is calculated from the observed temperature-speed
relationship at each spacecraft using the method of Lopez [1987]. We
did not require the two criteria to be satisfied everywhere in an
ICME interval. The boundaries of ICMEs are adjusted by incorporating
other signatures of ICMEs [Neugebauer and Goldstein, 1997, and
references therein]. A subset of ICMEs that have a strong magnetic
field, smooth field rotation and low proton temperature are defined
as magnetic clouds (MCs) [Burlaga et al., 1981]. The ICMEs are
further sorted into three categories: MCs and ICMEs (non-MCs) with
preceding shocks and ICMEs without preceding shocks. Events with
large data gaps in the sheath region are rejected, which gives a
total 18 MCs preceded by shocks, 21 ICMEs (non-MCs) with forward
shocks and 56 ICMEs without shocks between 1998 and 2005. All the
shocks upstream of the ICMEs are fast shocks across which the
magnetic field strength increases. We note that although
identification of ICMEs is a subjective art, the study of sheath
regions by definition requires a preceding shock. Since almost all
shocks inside 2 - 3 AU are generated by ICMEs, the uncertainties
inherent in identifying ICMEs are not a major difficulty for this
work.

\begin{figure}
\noindent\includegraphics[width=20pc]{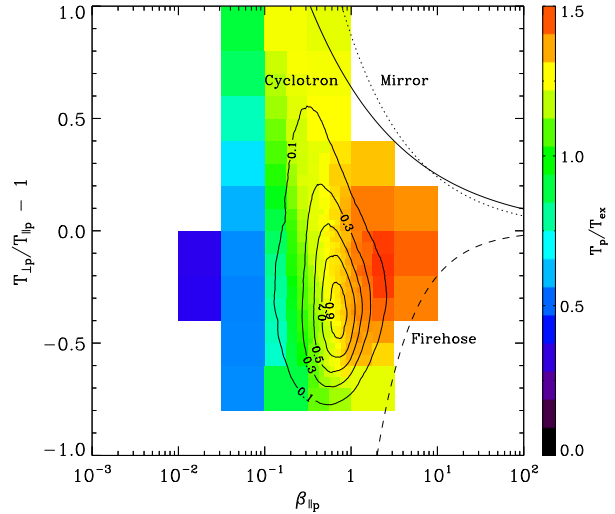} \caption{ACE
measurements of the normalized proton temperature $T_p/T_{ex}$ over
the thermal anisotropy and parallel plasma beta of protons. The
color shading indicates the average values of $T_p/T_{ex}$ for the
data binning. Black contours display the 2D histogram overlaid on
the data at levels of [0.1, 0.3, 0.5, 0.7, 0.9]. Also shown are the
thresholds for the firehose (dashed line), cyclotron (solid line)
and mirror (dotted line) instabilities.}
\end{figure}

We compare the thermal anisotropies determined from the Wind
non-linear fittings to those obtained from the ACE moment integrals
for different plasma regimes. Figure~2 shows ACE observations of the
normalized temperature $T_p/T_{ex}$ as a function of the temperature
anisotropy and parallel plasma beta of protons. The parallel plasma
beta of protons is defined as $\beta_{\parallel p} = {n_pk_{\rm
B}T_{\parallel p} \over B^2/2\mu_0}$. A similar plot for Wind data
can be found in Liu et al. [2006, Figure~9]. The $1.7\times 10^6$
ACE 64 s temperature averages from 1998 - 2005 are divided into
two-dimensional bins of temperature anisotropy and plasma beta; bins
with dense spectra are further subdivided, with the requirement that
each cell contains at least 2000 data points. The curves indicate
the threshold conditions
\begin{equation}
{T_{\perp p} \over T_{\parallel p}} - 1 = {S \over \beta_{\parallel
p}^{\alpha}}
\end{equation}
for the cyclotron, mirror and firehose instabilities, respectively.
The free parameters, $S$ and $\alpha$, are determined from solutions
to the single-fluid MHD equations or to the Vlasov dispersion
relation [Parker, 1958; Gary et al., 1997]. The firehose instability
has $S=-2$ and $\alpha=1$, while $S=0.64$, $\alpha=0.41$ for the
cyclotron instability over the range $0.1\leq \beta_{\parallel
p}\leq 10$ and $S=0.87$, $\alpha=0.56$ for the mirror mode in the
domain $5\leq \beta_{\parallel p}\leq 50$ at the maximum growth rate
$\gamma_m = 0.01\Omega_p$. Here $\Omega_p$ denotes the proton
cyclotron frequency. Most of the data are constrained by these
thresholds; when the thermal anisotropy is close to the thresholds,
the plasma is heated by the induced instabilities. As indicated by
the contour lines, the most probable temperature anisotropy for the
solar wind at 1 AU is $T_{\perp p}/T_{\parallel p}=$ 0.6 - 0.7. This
ratio can be increased in the sheath regions of fast ICMEs to 1.1 -
1.3. The ICME plasma, characterized by $T_p/T_{ex} \leq 0.5$, lies
far away from the instability thresholds. These results are
consistent with the picture described by Wind data [Liu et al.,
2006]. We conclude that the two methods used to derive the thermal
anisotropies give similar results.

\section{Case Study}
We first present a study of the association of a PDL and mirror mode
waves with an ICME observed by ACE and Wind on April 16 - 17 (day
106 - 107), 1999. Figure~3 shows the ACE data, with the ICME
interval indicated by the shaded area. The ICME was identified based
on the enhanced helium/proton density ratio, declining speed, and
depressed proton temperature; the smooth, strong magnetic field and
large rotation of the $B_y$ component (see Figure~5) indicate that
this event is an MC. A forward shock driven by the event passed the
spacecraft at 10:34 UT on April 16, with simultaneous sharp
increases in the proton density, bulk speed, temperature and
magnetic field strength. Within the sheath region, between the shock
and the leading edge of the MC, the proton density first increases
due to the compression at the shock and then decreases closer to the
MC. This density decrease is the plasma depletion mentioned above.
As shown by the bottom panel, fluctuations in the density and
magnetic field strength are out of phase during the density
increase; the correlation coefficient is about $-$0.6 for the data
subtracted by the backgrounds (refer to subsection~3.3) between
11:00 and 15:30 UT on April 16. The anti-correlation is also
revealed by Figure~6 combined with equations~(3) and (5). This
signature is typical for mirror waves. The second to last panel
shows the temperature anisotropy and instability threshold
conditions calculated from equation~2 for mirror and cyclotron
instabilities. The dotted line in this panel shows Wind data scaled
and time shifted to match the sheath and MC intervals seen at ACE.
Anisotropies derived from Wind fit and ACE moment integrals are in
agreement. The thermal anisotropy, ${T_{\perp p} \over T_{\parallel
p}}-1$, is near zero inside the MC but drops to $\sim -0.4$ in the
ambient solar wind. In the sheath region, the anisotropy is as high
as 0.5 (exceeding the instability thresholds) and has an average
value of 0.2 - 0.3 (very close to the thresholds). The plasma beta
(its effect is also reflected by the instability thresholds) is very
high in the sheath after 12:29 UT on April 16, with values sometimes
exceeding 50. As is evident from Figure~2, the mirror threshold is
lower than the cyclotron limit when $\beta_{\parallel p}>6$, which
is also indicated in Figure~3. Therefore, mirror instabilities are
excited in the sheath and give rise to the anti-correlated density
and magnetic field fluctuations. More evidence is provided below for
the plasma depletion and mirror mode structures.

\begin{figure}
\noindent\includegraphics[width=20pc]{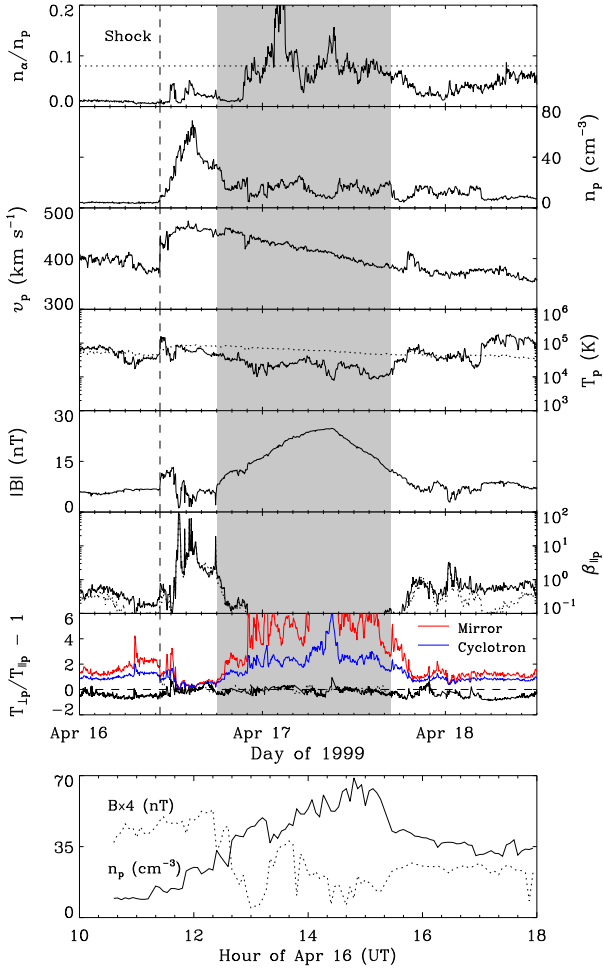} \caption{Solar wind
plasma and magnetic field parameters measured by ACE for a 2.5-day
interval in 1999. From top to bottom, the panels show the
alpha-to-proton density ratio, proton density, bulk speed, proton
temperature, magnetic field strength, parallel proton beta
$\beta_{\parallel p}$, thermal anisotropy, and an expanded view of
the magnetic and density fluctuations within the sheath. The shaded
region shows an MC. Dashed lines indicate the arrival time of the
MC-driven shock and the zero level of the anisotropy, respectively.
Dotted lines denote the 8\% level of the alpha/proton density ratio
(top panel), the expected proton temperature (fourth panel), the
perpendicular proton beta $\beta_{\perp p}$ (sixth panel), and the
scaled Wind data (seventh panel). Mirror and cyclotron thresholds
(computed from ACE data) are shown by the colored lines.}
\end{figure}

\subsection{Flux-rope and Shock Orientation}
The mechanism described in section 1 for generation of PDLs and
mirror mode waves may fit these observations. To test this fit
requires knowledge of the MC and shock orientation. Minimum variance
analysis (MVA) [Sonnerup and Cahill, 1967] provides estimates of the
MC axis orientation using the measured magnetic field vectors [e.g.,
Lepping et al., 1990; Bothmer and Schwenn, 1998]. Eigenvectors of
the magnetic variance matrix, $\hat{\bf y}^*$, $\hat{\bf z}^*$,
$\hat{\bf x}^*$, corresponding to the eigenvalues in descending
order, form a right-handed orthogonal coordinate system. MCs are
oriented along the intermediate variance direction ($\hat{\bf
z}^*$). Application of the MVA method to the normalized magnetic
field measurements within the MC gives an axis, in terms of the
azimuthal ($\theta$) and longitudinal ($\phi$) angles, of
$\theta=-69^{\circ}$ and $\phi=91^{\circ}$ in the geocentric solar
ecliptic (GSE) coordinates. We also determine the axis orientation
with a flux-rope reconstruction technique, based on the
Grad-Shafranov equation [Hau and Sonnerup, 1999; Hu and Sonnerup,
2002]; the resulting axis orientation is very close to the MVA
estimate.

Figure~4 is a 3D view of the MC modeled as a cylindrical flux rope
with a diameter of 0.23 AU (obtained from the average speed
multiplied by the MC duration in Figure~3) in the GSE system. Note
that the x axis points from the Earth to the Sun, so the Sun lies at
(1, 0, 0) AU in the system. The flux rope is viewed from (-2, 1.5,
0) AU relative to the origin (Earth) of the system and is almost
perpendicular to the ecliptic plane. The dark belt around the MC
represents part of the shock surface which is approximated as a
sphere centered at the Sun. The distance between the shock and the
MC is about 0.08 AU along the ACE trajectory. The upper arrow points
in the direction of the shock normal with $\theta=7^{\circ}$ and
$\phi=178^{\circ}$ determined from a least-squares fit to the
Rankine-Hugoniot relations [Vi\~{n}as and Scudder, 1986]. Magnetic
coplanarity gives a comparable result using the average magnetic
field vectors upstream and downstream of the shock. The angle
between the shock normal and upstream magnetic field is about
$61^{\circ}$, so the shock is oblique. The shock normal makes an
angle of about $96^{\circ}$ with the flux-rope axis. Therefore, the
whole situation can be pictured as a flux rope nearly perpendicular
to the ecliptic plane moving away from the Sun (along the -x
direction) and driving an oblique shock ahead of it. In the MC
frame, the ambient field lines may be draped around the MC,
analogous to planetary magnetosheaths.

\begin{figure}
\noindent\includegraphics[width=20pc]{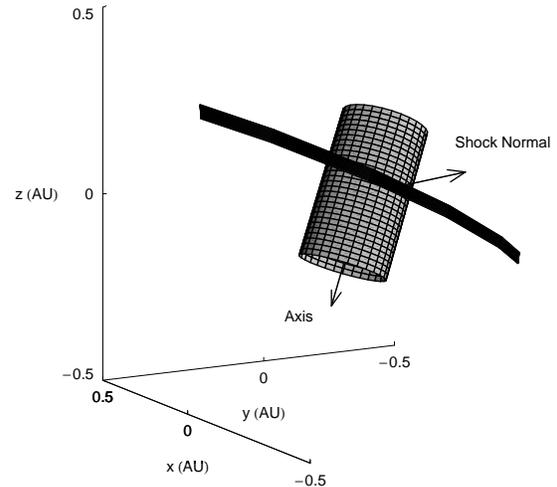} \caption{3D
rendering of the MC in GSE coordinates. Arrows indicate the axis
orientation and shock normal, respectively. The belt around the flux
tube approximates the shock surface.}
\end{figure}

The shock speed can be calculated from $v_s={n_2v_2-n_1v_1 \over
n_2-n_1}$ given by the conservation of mass across the shock.
Substituting the upstream density $n_1$ = 4 cm$^{-3}$ and velocity
$v_1$ = 375 km s$^{-1}$ and the downstream $n_2$ = 9 cm$^{-3}$ and
$v_2$ = 430 km s$^{-1}$, we obtain $v_s\simeq$ 474 km s$^{-1}$,
slightly larger than the MC leading-edge speed of 450 km s$^{-1}$.
The size of the sheath, between the MC and shock, is thus expected
to increase with distance. The MC is also expanding in the solar
wind, as shown by the speed difference between the leading and
trailing edges (see Figure~3). These two factors may affect the
formation of PDLs and anisotropic ion distributions within the
sheath.

\subsection{Plasma Depletion and Magnetic Field Draping}
Figure~5 provides an expanded view of the PDL and shows that it is
associated with field line draping. The dashed lines bracket the
PDL, in which the proton density decreases by a factor of 2 within
2.6 hours. Within the MC the $B_y$ component, not $B_z$ as usually
seen in MCs, exhibits a large rotation, which confirms our MC axis
determination. Interestingly, the magnetic field is nearly radial
inside this PDL as indicated by the comparison of the $B_x$
component with the field strength. Note that the spiral magnetic
field makes an angle of 45$^{\circ}$ with the radial direction on
average near the Earth; a radial field seldom occurs. The bottom
panel shows the dot product of the field direction with the axis
orientation, which is typically near zero in the PDL, so the field
lines in the PDL are perpendicular to the MC axis. This product
reaches its maximum close to the MC center. In contrast, the dot
product between the field direction and the minimum variance
direction ($\hat{\bf x}^*$) of the MC magnetic field is almost -1 in
the PDL but roughly constant throughout the MC interval. The
configurations of the flux rope and field lines within the PDL give
direct evidence for field line draping. The good time coincidence
between the density decrease and field line draping strengthens
confidence for these features.

\begin{figure}
\noindent\includegraphics[width=20pc]{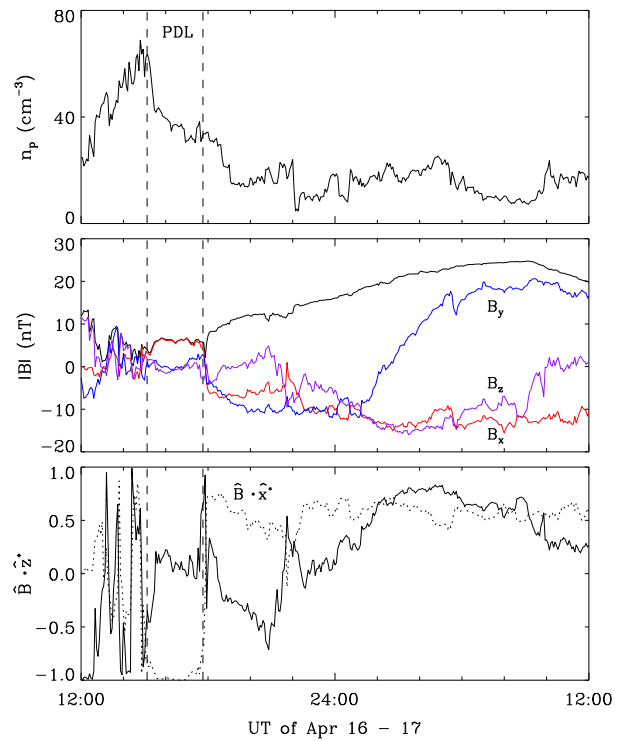} \caption{Proton
density (upper panel), magnetic field magnitude with its components
denoted by the colored lines (middle panel), and the dot product of
the field direction with the MC axis orientation (lower panel)
encompassing the PDL between the vertical dashed lines. The dotted
line in the bottom panel shows the dot product between the field
direction and the minimum variance direction of the MC magnetic
field.}
\end{figure}

\subsection{Wave Structures}
Mirror mode instabilities produce anti-correlated density and
magnetic field fluctuations, namely,
\begin{equation}
{\delta n \over n} = -({T_{\perp} \over T_{\parallel}}-1){\delta B
\over B}
\end{equation}
for a bi-Maxwellian plasma [e.g., Hasegawa, 1969], where $\delta n$
and $\delta B$ are perturbations in the background plasma density
$n$ and magnetic field strength $B$. However, in the same low
frequency regime the slow mode, alone among the three basic MHD
modes, also has a plasma density variation out of phase with the
magnetic field fluctuation. Starting with the CGL theory [Chew et
al., 1956], we obtain the relationship between density and magnetic
field fluctuations
\begin{equation}
{\delta n \over n} = ({P_{\perp}\sin^2\theta \over
nm_p})^{-1}[{\omega^2 \over k^2}-{B^2 \over \mu_0nm_p} - {P_{\perp}
\over nm_p} + {P_{\parallel}\cos^2\theta \over nm_p}]{\delta B \over
B},
\end{equation}
where $m_p$ is the proton mass, $\theta$ is the angle between the
wave vector ${\bf k}$ and the magnetic field, and $P_{\perp}$,
$P_{\parallel}$ are the perpendicular and parallel thermal
pressures. The dispersion relation, ${\omega^2 \over k^2}$, is given
by Parks [2004, equation (9.247), see the errata]. Equation (4) is a
general relationship between the density and magnetic fluctuations
for the slow and fast modes. It can be shown from equation (4) that
the fast mode always gives positively correlated $\delta n$ and
$\delta B$. The firehose instability arising from this mode will
also induce in-phase density and magnetic field fluctuations, at
least in its linear stage. The intermediate mode with a dispersion
relation ${\omega^2 \over k^2}=v_{\rm A}^2({P_{\perp}-P_{\parallel}
\over B^2/\mu_0}+1)\cos^2\theta$ is not involved with the
perturbations so that $\delta n=\delta B = 0$ across this mode as in
the ordinary MHD theory. Here $v_{\rm A}$ denotes the Alfv\'{e}n
speed.

The direction of the wave vector can be estimated using  MVA on the
measured magnetic field. For plane waves, the propagation vector
lies along the minimum variance direction. The best estimate of the
direction is roughly normal to the magnetic field with an average
value of 93$^{\circ}$ for the interval with active fluctuations in
the sheath region, so the waves may be highly oblique. Therefore,
equation (4) can be reduced to
\begin{equation}
{\delta n \over n} = -(1+{2\over \beta_{\perp}}){\delta B \over B}
\end{equation}
for the slow mode. Equations (3) and (5) can be used to compare data
with theoretical predictions.

Figure~6 displays the density variations within a 4-hour interval in
the sheath region. Note that the fluctuations seem to be periodic at
later times. A frequency of $1.04\times 10^{-3}$ Hz stands out in
the power spectrum of the fluctuations, which corresponds to a
period of about 960 s. The smooth density profile is obtained by
applying a Butterworth low-pass filter to the frequency space of the
data and then converting it back to the time domain. We set the
cutoff frequency of the filter to $5\times 10^{-4}$ Hz, smaller than
the present wave frequency. The same filter is also applied to the
magnetic field data to get the background field strength. The filter
removes fluctuations with frequencies higher than the cutoff
frequency, so the smoothed quantities represent the unperturbed
profiles. Using these background quantities and the field
variations, we obtain the density perturbations predicted by the
mirror and slow modes from equations (3) and (5), respectively.
Figure~6 compares the observed and predicted density fluctuations.
The density variations predicted by the mirror mode agree fairly
well with observations after 12 UT on April 16; only one spike is
missed at 14:17 UT. The slow mode also predicts the phase of the
variation correctly, but gives much larger fluctuation amplitudes
than observed. As implied by equation (3), the mirror mode density
perturbation relative to the background is smaller in amplitude than
the corresponding magnetic field variation for a moderate
anisotropy, while the opposite is true across slow mode waves as
described by equation (5). Stasiewicz [2004] reinterpreted the
mirror mode fluctuations in the terrestrial magnetosheath as trains
of slow mode solitons. In order to suppress the density variations,
the author adopts $\kappa=0$, $\gamma=1.7$ for the relationship
$P_{\perp} \sim n^{\gamma}B^{\kappa}$, whereas the CGL theory gives
$\kappa=1$, $\gamma=1$ as required by the conservation of the
magnetic moment. Therefore, we conclude that the wave activity is
driven by the mirror mode. The density and magnetic field
fluctuations in the PDL are significantly reduced (see Figure~3),
probably due to the decrease in the plasma beta.

\begin{figure}
\noindent\includegraphics[width=20pc]{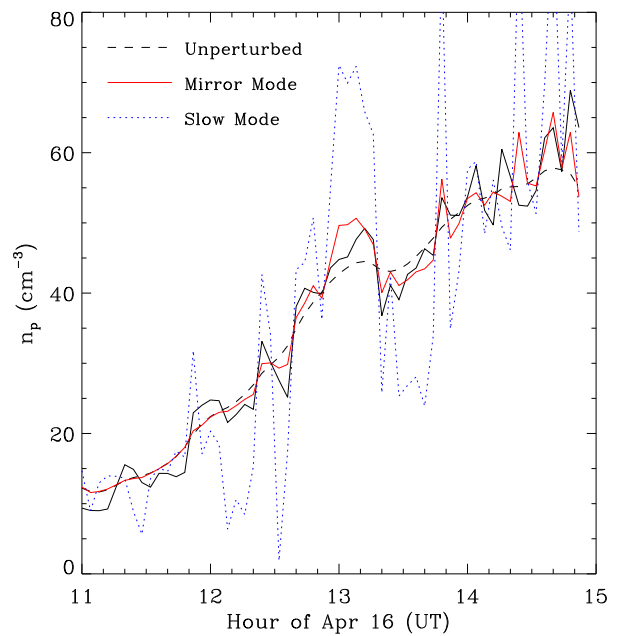} \caption{Expanded
view of the density fluctuations ahead of the PDL. Colored lines
indicate density variations predicted by the slow and mirror modes,
respectively. The background profile obtained from a low-pass filter
is represented by the dashed line.}
\end{figure}

Note that the mirror mode is non-oscillatory and its proper
treatment requires kinetic theory [e.g., Hasegawa, 1969; Southwood
and Kivelson, 1993]. It arises as the result of the mirror
instability which is a purely growing mode in a uniform plasma. The
mirror mode may couple to propagating waves, for instance, the drift
wave produced by magnetic field and density gradients [Hasegawa,
1969], and then convect away. The oscillatory structure in Figure~6
may result from this wave coupling. To illustrate this point, we
write the dispersion relation as [Hasegawa, 1969]
\begin{equation}
\omega = \omega^{\ast}+i\xi\mu^{1/2}[\eta-(1+{\mu\over \nu}){e^{\nu}
\over I_0(\nu)-I_1(\nu)}],
\end{equation}
where
\begin{eqnarray}
\hspace{2cm}\xi=({2\over \pi})^{1/2}({T_{\parallel} \over
T_{\perp}})^{3/2}{\Omega_p\over \beta_{\perp}}, \nonumber \\
\eta = \beta_{\perp}({T_{\perp}\over T_{\parallel}}-1),
\nonumber\\
\mu = {k_{\parallel}^2\over \Omega_p^2} {k_{\rm
B}T_{\perp} \over m_p}, \nonumber \\
\nu = {k_{\perp}^2\over \Omega_p^2} {k_{\rm B}T_{\perp} \over m_p}.
\nonumber
\end{eqnarray}
Here $\omega^{\ast}$ is the frequency of the wave that the mirror
mode is coupled to, and $I_0$, $I_1$ are the modified Bessel
functions of the first kind with order 0 and 1. Note that the
electron contribution to this dispersion relation is factored out by
assuming a cold electron distribution. Without the wave coupling,
the wave frequency $\omega$ is purely imaginary and as a result the
mirror wave is non-propagating. This point contradicts the
perspective of fluid theory. The onset condition expressed by
equation (1) is derived from the above equation in the long
wavelength limit when the perpendicular wave number $k_{\perp}$ is
much larger than the parallel wave number $k_{\parallel}$. Here we
use equation (6) to estimate the maximum growth rate of the mirror
instability. Shown by this equation, the growth rate is proportional
to $k_{\parallel}$. While $k_{\perp}\gg k_{\parallel}$, the maximum
growth rate will be restricted by the effect of finite Larmor radius
since the perpendicular wavelength cannot be smaller than the ion
Larmor radius. Taking derivatives of that equation with respect to
$\mu$ and $\nu$ and equating them to zero, we obtain the maximum
growth rate expressed as
\begin{equation}
\gamma_m = ({3\over \pi})^{1/2}{1\over
6\beta_{\perp}}({T_{\perp}\over
T_{\parallel}})^{-3/2}(\eta-1)^2\Omega_p
\end{equation}
for $k_{\parallel}\ll k_{\perp}$ in the long wavelength limit. Note
that $\eta$ is a measure of the overshoot of the mirror instability
which must be larger than 1 as required by equation (1). The
theoretical growth rate is displayed in Figure~7 over the two
dimensional plane of the parallel plasma beta and thermal anisotropy
under the constraint of $\eta > 1$. As is evident from the figure, a
significant growth rate occurs only in high beta regions when the
thermal anisotropy is moderate. The best power law fit to the
$0.01\Omega_p$ level of the growth rate over the range $5\leq
\beta_{\parallel p}\leq 50$ in the form of equation (2) gives
$S=1.05$ and $\alpha=0.85$, close to but lower than the the
numerical evaluation of Gary et al. [1997] in the same domain. In
the current case, the average observed values $\beta_{\perp p}\simeq
12$ and ${T_{\perp p}\over T_{\parallel p}} \simeq 1.2$ result in
$\gamma_m = 0.02 \Omega_p$, which seems reasonable for the overshoot
of the instability of $\eta\simeq 2.4$.

\begin{figure}
\noindent\includegraphics[width=20pc]{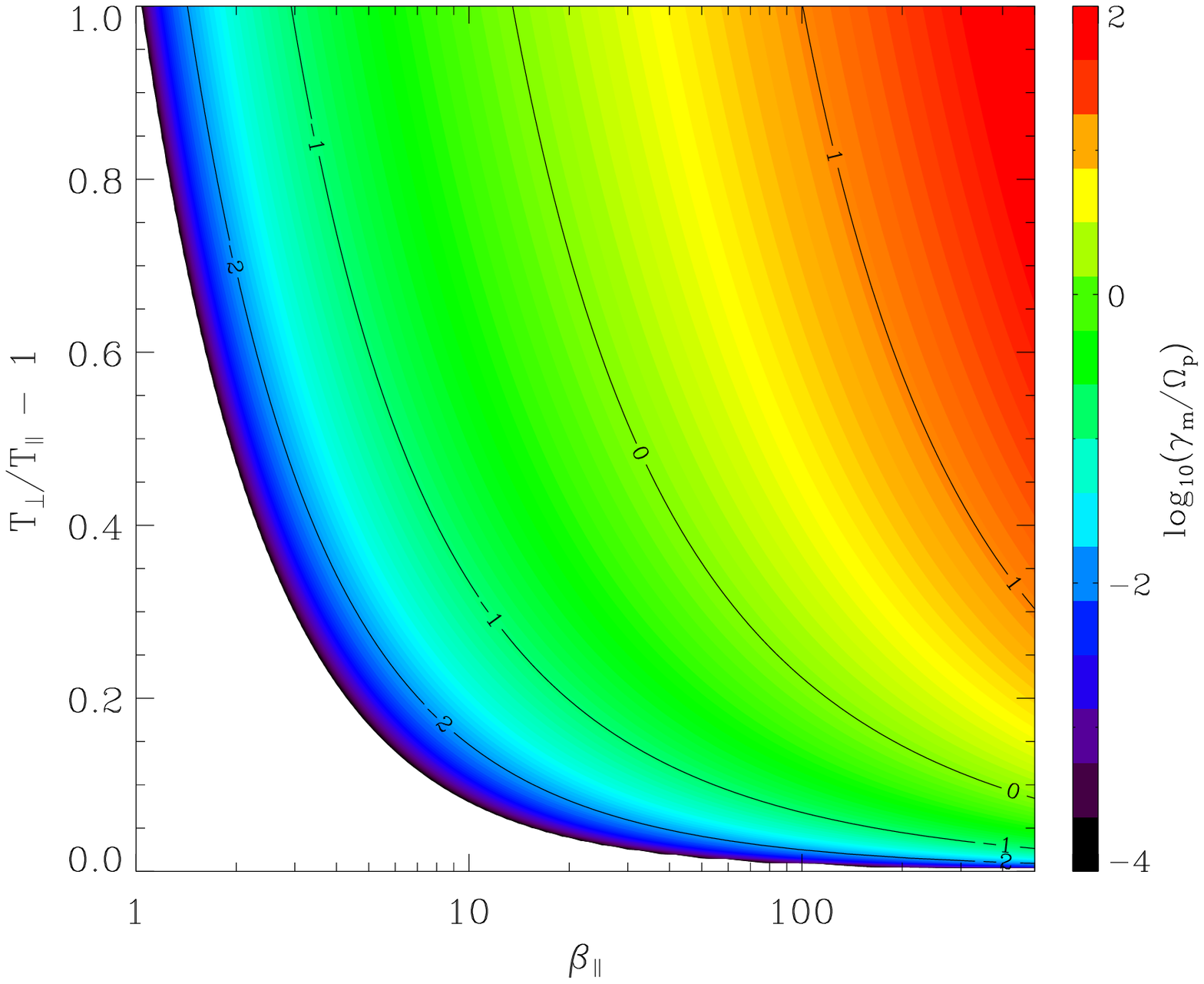} \caption{Theoretical
growth rate of the mirror instability as a function of the parallel
plasma beta and temperature anisotropy. The color bar indicates the
logarithmic scale of the growth rate in units of the proton
cyclotron frequency. Contour lines show the growth rate levels of
[$10^{-2}, 10^{-1}, 10^0, 10^1]\times \Omega_p$.}
\end{figure}

\section{Superposed Epoch Analysis}
In the previous section, we presented an example showing the
presence of a PDL and mirror mode waves in front of an MC. ICME data
in the literature often show these features [e.g., Neugebauer and
Goldstein, 1997, Figure~3; Bothmer and Schwenn, 1998, Figures~4, 5,
6, 7 and 8; Webb et al., 2000, Figure~6; Mulligan and Russell, 2001,
Figures~1 and 2; Richardson et al., 2002, Figure~2; Richardson et
al., 2004, Figure~1; Zurbuchen and Richardson, 2004, Figure~2;
Gosling et al., 2005, Figure~1]; all these ICMEs are preceded by
shocks. In this section, we use a superposed epoch analysis (SEA) to
give a broad-based view of these features. We define the ICME
arrival time as the zero epoch and superpose the ICME-related data
for the three ICME classes (MCs with shocks, non-MC ICMEs with
shocks, ICMEs without shocks) separately. Typical uncertainties in
identifying the beginning of an ICME are estimated to be 1 - 2 hours
depending on the time resolution of the data. ICMEs with preceding
shocks are scaled into a 30 hour long interval (the average duration
of an ICME at 1 AU) and the sheath regions are scaled into 14 hour
intervals (the average duration of the sheath at 1 AU); ICMEs
without shocks are scaled into the same interval, but since there is
no sheath region we use the 14 hours of data ahead of the ICMEs.
Thus we line up the data in a fixed time relation to the ICME
arrival times as if we had many observations of a single event. By
averaging the superposed data for each time, real signals will be
preserved but noise will tend to average out [e.g., Haurwitz and
Brier, 1981; Prager and Hoenig, 1989].

\begin{figure*}
\centerline{\includegraphics[width=30pc]{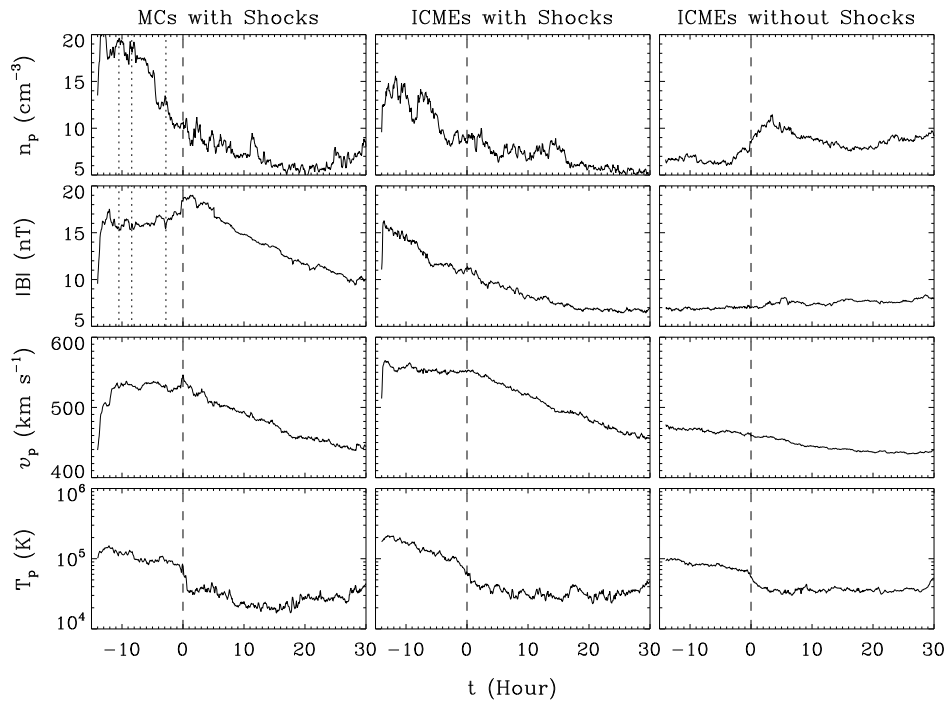}}
\caption{Superposed epoch plots of the proton density, magnetic
field strength, bulk speed and proton temperature for the three
classes of ICMEs. The zero time (dashed lines) is the ICME forward
boundary. Dotted lines align some magnetic field dips with the
corresponding density spikes in the MC sheath.} \end{figure*}

Figure~8 displays the superposed 4 m averages of ACE plasma and
magnetic data for the three categories of ICMEs. As required by our
ICME identification criteria, the ICME intervals (the data after the
zero epoch) have a depressed proton temperature and the MCs (left
panels) possess a strong magnetic field. The first two classes of
ICMEs have expansion speeds (the speed difference between the
leading and trailing edges) of about 90 km s$^{-1}$; the third class
of ICMEs, those without preceding shocks, has an expansion speed of
about 20 km s$^{-1}$. As a result of this expansion, the first two
ICME classes have a lower average plasma density than the third
class. Using the ambient solar wind density $n=6.5$ cm$^{-3}$,
temperature $T=10^5$ K and magnetic field strength $B=7$ nT derived
from the data ahead of the ICMEs in the right panels, we estimate
the fast mode speed from $v _f= (v_{\rm A}^2 + {5k_{\rm B}T \over
3m_p})^{1/2}$ to be around 70 km s$^{-1}$. The expansion speed can
be regarded as the ICME speed relative to the ambient solar wind,
which is larger than the fast mode speed for the first two ICME
classes. Consequently, shock waves should be driven ahead of the
fast ICMEs, consistent with the data in Figure~8. For all ICMEs, the
expansion speed is of order of the Alfv\'{e}n speed, 50 - 60 km
s$^{-1}$  at 1 AU. This result is in agreement with previous
findings [e.g., Burlaga et al., 1981; Liu et al., 2005].

In the left and middle panels, the superposed shock occurs 14 hours
before the arrival time of the ICMEs; between the shock and the ICME
is the sheath region where PDLs and mirror mode waves may occur. For
MCs with preceding shocks (left panels), a PDL is present beginning
about six hours ahead of the ICME in which the proton density
decreases by a factor of about 1.7 and the field magnitude gradually
increases. The anti-correlated density and magnetic field
fluctuations, with some times marked by the dotted lines, show that
wave structures are also present. Presumably these structures are
induced by the mirror instability; the superposed thermal anisotropy
supports this interpretation (see Figure~9). Plasma depletion may
also occur in the sheath region of ICMEs with shocks (middle
panels). The plasma is rarefied in the layer adjacent to these
ICMEs, but the magnetic field magnitude remains roughly constant.
The density and magnetic field have high-frequency fluctuations, but
no coherent structure is apparent. The sheath regions of the first
two ICME classes thus show a different association with the PDL and
mirror waves, which may reveal the dependence of these features on
the ICME geometry. MCs are better organized than common ICMEs in
terms of magnetic structure and may lead to favorable conditions for
such features to develop. The ICMEs not associated with shocks
(right panels) do not have PDLs and the density and magnetic
fluctuations are even further reduced. The compression of the
ambient solar wind by these ICMEs is apparently not large enough to
produce these features, contrary to the suggestion of Farrugia et
al. [1997] that a PDL may form irrespective of whether ICMEs drive
shocks or not. Figure~8 indicates that shocks may be necessary to
produce these features, since their formation is closely related to
the extent of the compression. Observations of PDLs in planetary
magnetosheaths show a gradual increase in the magnetic field
simultaneous with the density decrease [e.g., Crooker et al., 1979;
Hammond et al., 1993; Violante et al., 1995]. This increase in the
field strength may not be required in ICME sheaths. The upstream
field compression by ICMEs may be alleviated by the sheath expansion
as the preceding shocks move away from the ejecta.

The speed and field strength profiles within ICMEs are smooth
compared with the turbulent profiles of the density and temperature.
As is well known, Alfv\'{e}n waves give rise to correlated velocity
and magnetic field fluctuations [e.g., Stix, 1962]. The
low-frequency Alfv\'{e}n waves may be responsible for solar wind
heating and acceleration through a non-linear cascade process of
energy [e.g., Isenberg and Hollweg, 1983; Goldstein et al., 1995;
Leamon et al., 1999; Hu and Habbal, 1999]. Radial evolution of ICMEs
is investigated by Liu et al. [2006]; they find that the ICME
temperature decreases more slowly with distance than an adiabatic
profile. This plasma heating may not be powered by Alfv\'{e}n waves
since they are not prevalent within ICMEs as implied by the smooth
speed and magnetic field profiles.

\begin{figure*}
\centerline{\includegraphics[width=30pc]{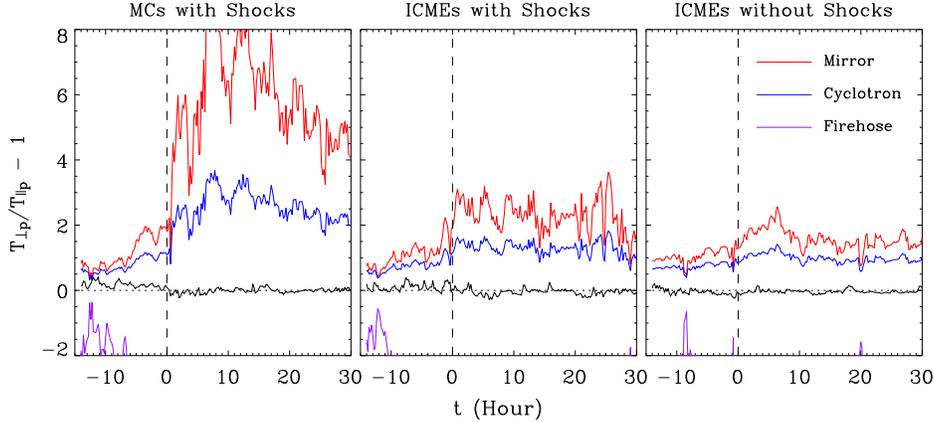}}
\caption{Superposed epoch plots of proton thermal anisotropy for the
same three classes of ICMEs as in Figure~8. Dashed lines mark the
ICME arrival times while dotted lines show the zero level of the
anisotropy. Threshold conditions for the instabilities are
represented by the colored lines.}
\end{figure*}

Mirror waves can also be evidenced by the temperature anisotropy.
Figure~9 shows the superposed thermal anisotropy data from Wind for
the three classes of ICMEs. The threshold instability conditions,
calculated from the superposed parallel plasma beta
$\beta_{\parallel p}$ using equation (2), are plotted to allow
comparison with the anisotropy. An abrupt increase in the mirror and
cyclotron thresholds can be seen at the leading edge of the first
two groups of ICMEs (left and middle panels); it is stronger in the
first case, indicating that the plasma beta is low within these MCs.
For the third ICME class (right panel), the onset conditions inside
and outside the ICMEs are similar. The ICME plasma typically has an
anisotropy near zero, so it is stable with respect to temperature
anisotropy instabilities. This point is consistent with the picture
described by Figure~2. The thermal anisotropy increases to 0.1 - 0.4
in the sheath regions of the first ICME category (left panel) and is
very close to the instability thresholds at some epochs. The sheath
plasma may be marginally unstable to the mirror and cyclotron
instabilities. As a result, mirror waves may be induced under
certain conditions and then be carried away by coupling to other
propagating modes. This interpretation is in accord with the
anti-correlated density and magnetic field fluctuations seen in
Figure~8 (left panels). The temperature anisotropy may also be
enhanced in the sheaths of the second ICME class (middle panel), but
not as strongly, so the fluctuations are reduced as shown in
Figure~8 (middle panels). For the third ICME group (right panel),
the thermal anisotropy does not deviate from zero across the sheath
and ICME intervals, so mirror waves do not appear.

The superposed parallel proton beta $\beta_{\parallel p}$ ranges
from 0.5 to 5.5 in the sheath regions of the MCs with preceding
shocks. In this regime, the instabilities are dominated by the
cyclotron mode since it has a lower threshold than the mirror mode
(refer to Figure~2). Gary [1992] showed through numerical evaluation
of the Vlasov dispersion relation that the cyclotron instability has
a higher linear growth rate. The temperature anisotropy can be
rapidly reduced by the cyclotron instability and thus the mirror
mode may not develop to a significant level. However, Price et al.
[1986] found that the introduction of a small fraction of minor ions
would substantially lower the growth rate of the cyclotron
instability while leaving the mirror mode unaffected. Since helium
ions are always present in the solar wind, the mirror mode would be
able to grow given the observed plasma beta and thermal anisotropy.

It should be noted that the SEA smoothes the data. Consequently, the
superposed field magnitude and density do not exhibit large
fluctuations. The smoothed profiles of the threshold conditions and
thermal anisotropy also underestimate the marginal instability to
the mirror mode. Examining ICMEs individually is the best way to see
the mirror mode structures, as demonstrated by the case study in
section~3. The references listed in the first paragraph of section~4
give ICMEs that often show the mirror mode fluctuations (and PDLs)
in their sheath regions; large dips in the magnetic field strength
serve as a good identifier for the mirror mode structures.

\section{Conclusions and Discussion}
In analogy with planetary magnetosheaths, we propose that PDLs and
mirror mode structures form in the sheath regions of fast ICMEs. The
upstream field compression by fast ICMEs and their leading shocks
results in field line draping and anisotropic plasma heating in the
sheath. An extensive case study and statistical analysis of ICMEs
observed by ACE and Wind show that these two features occur in the
sheaths ahead of fast events.

The association of plasma depletion and mirror waves with ICMEs is
demonstrated by an MC example observed at ACE and Wind on 16 April
1999. This event drives a forward oblique shock with its axis
perpendicular to the shock normal. Downstream of the shock, the
proton density decreases by a factor of 2 within 2.6 hours
(corresponding to 0.03 AU in length) ahead of the MC. At the same
time the  magnetic field is stretched to be nearly radial and normal
to the MC axis. Anti-correlated density and field strength
fluctuations are seen inside the sheath region between the MC and
the shock and are consistent with mirror mode waves; the thermal
anisotropy exceeds the mirror mode onset condition and the plasma
beta is high. Analytical growth rates of the mirror instability
limited by the effect of finite Larmor radius are obtained as a
function of plasma beta and temperature anisotropy. From the
observed overshoot of the mirror instability the maximum growth rate
is estimated to be around $0.02\Omega_p$.

We perform a SEA on three classes of ICMEs to investigate the
general properties of PDLs and mirror mode waves associated with
these different kinds of events. On average, the sheath region is
about 14 hours long (roughly 0.17 AU) at 1 AU in comparison with the
0.2 - 0.3 AU span of ICME intervals. For MCs preceded by shocks,
PDLs are observed to have an average density decrease of a factor of
1.7 which lasts about 6 hours and is accompanied by a gradual
increase in the field magnitude. The average thermal anisotropy
${T_{\perp p}\over T_{\parallel p}} \simeq$ 1.2 - 1.3, close to the
threshold for the mirror instability, leads to anti-correlated
fluctuations in the density and magnetic field strength. Compared
with the MCs, non-MC ICMEs with forward shocks have a thinner plasma
depletion layer close to their leading edges. The thermal anisotropy
is only slightly enhanced, so fluctuations in the density and
magnetic field are smaller and do not have a definite structure.
This difference between the two classes may indicate the effect of
the ICME geometry in creating the features. The third category,
ICMEs without shocks, is not associated with plasma depletion and
mirror waves. The occurrence of these signatures may be determined
by the extent of the upstream field compression by ICMEs. As a
measure of the compression, an ICME speed of about 90 km s$^{-1}$
relative to the ambient solar wind seems necessary to drive shocks
and produce the features. As revealed by the SEA, all the ICME
plasma is stable to the temperature anisotropy instabilities,
consistent with the finding of Liu et al. [2006].

These results reveal important physical processes which alter the
ICME environment and provide another setting in which to study PDLs
and mirror mode waves. As noted by Zwan and Wolf [1976], PDLs can
only develop in the absence of significant magnetic reconnection;
otherwise, flux tubes that are compressed against the magnetosphere
would merge with geomagnetic field lines before they are diverted
around the obstacle. The same condition should hold in ICME sheaths
for the squeezing process to be operative. An ideal MHD simulation
of plasma flow behind MC-driven shocks also seems to confirm that
the PDL becomes thinner with a small increase in the reconnection
rate [Erkaev et al., 1995]. McComas et al. [1988; 1994] suggest that
magnetic reconnection should commonly take place between fast ICMEs
and the upstream ambient solar wind in the same manner as occurs at
the dayside of the magnetopause. Direct evidence for the local
reconnection in the solar wind was not provided until recently, but
none of the reconnection sites are at the interface between ICMEs
and the upstream solar wind [Gosling et al., 2005]. Given the
frequency of PDLs in ICME sheaths, magnetic reconnection may not be
prevalent or locally important in the sheath regions of fast ICMEs.

Observations of planetary magnetosheaths indicate that mirror waves
can make large depressions, i.e., holes in the background magnetic
field [e.g., Kaufmann, et al., 1970; Tsurutani et al., 1992;
Violante et al., 1995]. This point is emphasized by Winterhalter et
al. [1994] who made a survey of magnetic holes observed at Ulysses
and examined their relationship with mirror instabilities.
Consistent with the results of Klein and Burlaga [1980], they find
that the holes tend to occur in the interaction regions where fast
streams overtake the ambient solar wind and the mirror mode is
marginally stable. Non-linear saturation mechanisms of the mirror
instability are qualitatively discussed by Kivelson and Southwood
[1996]. In the non-linear saturation process, marginal stability can
be achieved by large reductions in the magnetic field, so the fully
evolved state would be characterized by holes in the background
magnetic field rather than alternate field enhancements and
depressions. However, Bavassano Cattaneo et al. [1998] suggest that
mirror waves make field enhancements as well as dips based on
Voyager observations of Saturn's magnetosheath. According to
Figure~7, the mirror instability will grow quickly and consequently
make a series of holes when the plasma beta is high; adjacent holes
are so close that the magnetic field appears to be alternately
enhanced and depressed. When the plasma beta is occasionally high,
the hole will be isolated. Their observed plasma beta profile seems
to support this explanation. If the interpretation that magnetic
holes are remnants of mirror mode structures is correct, the
mechanism suggested here would be able to explain the creation of
some magnetic holes localized in the solar wind in a self-consistent
manner.

Mirror waves may contribute to energetic particle modulation and
acceleration in the sheath regions of ICMEs. Various studies of the
cosmic ray modulation in the solar wind show that enhanced magnetic
turbulence in the sheath is particularly effective in producing
Forbush decreases [e.g., Badruddin et al., 1991; Ifedili, 2004],
probably due to particle scattering by waves or their non-linear
evolved states. According to Kivelson and Southwood [1996], the
final evolved state of mirror waves would be such that the total
perpendicular pressure (plasma plus field) is constant along the
field line. Particles with small pitch angles will be constrained by
the mirror force $F=-\mu\nabla_{\parallel}B$, where $\mu$ is the
magnetic moment; if the magnetic moment and total energy are
invariant, motion of these particles into weak field regions along
the field line will convert perpendicular energy to parallel and
kinetic energy, which can serve to suppress growth of the mirror
instability. Particles with large pitch angles may be excluded from
strong field regions, also leading to the pressure balance.

\appendix
\section{Test of Superposed Epoch Analysis}
Superposed epoch analysis (SEA) is useful to identify the
association of individual features with key events (here defined as
the ICME arrival times) in time series data. The statistical
significance of the association can be determined by a randomization
technique which avoids the assumptions of normality, random sampling
and sample independence made by parametric testing [e.g., Haurwitz
and Brier, 1981; Prager and Hoenig, 1989]. The null hypothesis is
that plasma depletion and mirror waves are not associated with
ICMEs. Under this null hypothesis, any time in the data can be
considered as a key event, i.e., an ICME arrival time.

We use the 4 m averages of ACE proton density data from 1998 - 2005
as a proxy for the significance test. We repeatedly draw random sets
of locations for the key event and their corresponding background
times from the data. Key events closer than a minimum spacing of 27
days are discarded to avoid the effect of solar sector crossings.
Each randomization yields 39 positions of simulated ICME arrival
times and 660 background spectra corresponding to each key event
(210 before and 450 after). For each set, we compute the t-statistic
defined as
$$t={\bar{E}-\bar{B} \over \sqrt{{(m+n)(m\sigma_E^2+n\sigma_B^2)
\over mn(m+n-2)}}},$$ where $\sigma_E$ and $\sigma_B$ are the
standard deviations of the $m=39$ key events and $n=660\times 39$
background times respectively, and $\bar{E}$, $\bar{B}$ are their
corresponding means. This statistic obeys Student's t-distribution
with degrees of freedom $m+n-2$ for normally distributed and
independent data. The observed t-statistic is calculated from the
true arrival times of the 39 ICMEs with preceding shocks used in our
analysis. The frequency distribution of the simulated t-statistic
resulting from 10,000 runs of the randomization is displayed in
Figure~10 as a histogram with a bin size of 0.1. Compared with the
Gaussian fit, the distribution skews toward the high-value tail. The
significance level can be estimated from the relative frequency with
which the simulated t-statistic is smaller than or equal to the
observed one. The observed statistic is about 0.91, corresponding to
a significance level of 91\%. In contrast, the standard
t-distribution gives 82\% for the observed statistic.

\begin{figure}
\centerline{\includegraphics[width=20pc]{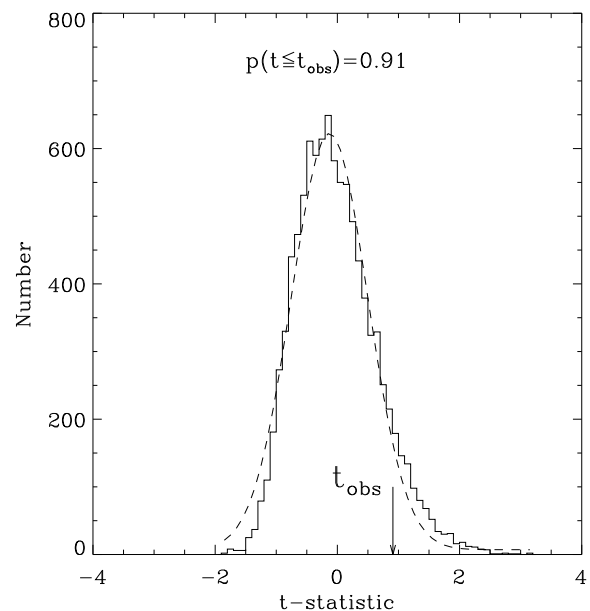}}
\caption{Histogram distribution of the t-statistic from the Monte
Carlo simulations with its Gaussian fit denoted by the dashed line.
The arrow indicates the value of the real statistic while the upper
text shows the significance level.}
\end{figure}

%
%

\begin{acknowledgments}
We acknowledge the use of the magnetic field data of ACE and Wind
from the NSSDC. The work at MIT was supported under NASA contract
959203 from JPL to MIT, NASA grants NAG5-11623 and NNG05GB44G, and
by NSF grants ATM-0203723 and ATM-0207775.
\end{acknowledgments}

%
%

%
%
%

%
%

\end{article}

\end{document}